%% file: pmain.tex
\newif\ifcomment
\newif\ifdraft
\newif\iflatexdiff
\latexdifftrue

\def\dvers{v7}
\def\dtitle{Analysis of the apparent nuclear modification\\ in peripheral \PbPb\ collisions at 5.02 TeV}
\def\stitle{\dtitle} 
\documentclass[ALICE,manyauthors,12pt]{cernphprep}
\usepackage[comma,square,numbers,sort&compress]{natbib}
\usepackage{hyperref}
\usepackage{graphicx}  
\usepackage{dcolumn}   
\usepackage{bm}        
\usepackage{amssymb}   
\usepackage{amsfonts}
\usepackage{graphics}
\usepackage{grffile}   
\usepackage{epsfig}
\usepackage[loose]{units}
\usepackage[usenames,dvipsnames]{xcolor}
\usepackage[normalem]{ulem} 
\usepackage{color}
\usepackage[utf8]{inputenc}
\usepackage[T1]{fontenc}
\usepackage{subfigure}
\iflatexdiff
\RequirePackage{color}\definecolor{RED}{rgb}{1,0,0}\definecolor{BLUE}{rgb}{0,0,1}

\fi

\input{commands.tex}
\graphicspath{{./img/}}
\ifdraft
\usepackage{lineno}
\modulolinenumbers[1]
\usepackage{fancyhdr}
\fi
\begin{document}
\begin{titlepage}
\PHyear{2018}
\PHnumber{115} 
\PHdate{9 May}   
\title{\dtitle}
\ShortTitle{\stitle}
\Collaboration{ALICE Collaboration%
         \thanks{See Appendix~\ref{app:collab} for the list of collaboration members}}
\ShortAuthor{ALICE Collaboration} 
\begin{center}
\ifdraft
\today\\ \color{red}DRAFT \dvers\ \hspace{0.3cm} \$Revision: 5032 $\color{white}:$\$\color{black}\vspace{0.3cm}
\else
\today
\fi
\end{center}
\begin{abstract}
Charged-particle spectra at midrapidity are measured in Pb--Pb collisions at the centre-of-mass energy per nucleon--nucleon pair $\snn = 5.02$ TeV and presented in centrality classes ranging from most central~(0--5\%) to most peripheral~(95--100\%) collisions.
Possible medium effects are quantified using the nuclear modification factor~($\RAA$) by comparing the measured spectra with those from proton--proton collisions, scaled by the number of independent 
nucleon--nucleon collisions obtained from a Glauber model.
At large transverse momenta~($8<\pT<20$~\gevc), the average $\RAA$ is found to increase from about $0.15$ in 0--5\% central to a maximum value of about $0.8$ in 75--85\% peripheral collisions, beyond which it falls off strongly to below $0.2$ for the most peripheral collisions. 
Furthermore, \RAA\ initially exhibits a positive slope as a function of $\pT$ in the $8$--$20$~\gevc\ interval, while for collisions beyond the 80\% class the slope is negative. 
To reduce uncertainties related to event selection and normalization, we also provide the ratio of $\RAA$ in adjacent centrality intervals.
Our results in peripheral collisions are consistent with a PYTHIA-based model without nuclear modification, demonstrating that biases caused by the event selection and collision geometry can lead to the apparent suppression in peripheral collisions.
This explains the unintuitive observation that \RAA\ is below unity in peripheral \PbPb, but equal to unity in minimum-bias \pPb\ collisions despite similar charged-particle multiplicities.
\end{abstract}
\end{titlepage}
\newpage
\setcounter{page}{2}
\section{Introduction}
\label{sec:intro}
Transport properties of the Quark-Gluon Plasma (QGP) can be extracted from measurements of observables in high-energy nucleus--nucleus~(\AaAa) collisions, which involve large momentum transfers, such as \com{high transverse momentum~($\pt$) }jets originating from hard parton-parton scatterings in the early stage of the collision. 
While propagating through the expanding medium, these hard partons lose energy due to medium-induced gluon radiation and collisional energy loss, a process known as ``jet quenching''~\cite{Gyulassy:1990ye,Baier:1994bd}.
Due to the energy loss, the rate of high-$\pt$ particles is expected to be suppressed relative to proton--proton collisions.
The effect is typically quantified by the nuclear modification factor 
\begin{equation}
\RAA = \frac{1}{\avNcoll} \, \frac{{\rm d}N_{{\rm ch}}^{{\rm AA}}/{\rm d}\pT}{{\rm d}N_{\rm ch}^{{\rm pp}}/{\rm d}\pT}\, =  \frac{1}{\avTaa} \, \frac{{\rm d}N_{{\rm ch}}^{{\rm AA}}/{\rm d}\pT}{{\rm d}\sigma_{\rm ch}^{{\rm pp}}/{\rm d}\pT}\, \,,
\label{eq:raa}
\end{equation}
defined as the ratio of the per-event yields in \AaAa\ and \pp\ collisions normalized to an incoherent superposition of \avNcoll\ binary \pp\ collisions. 
The average number of collisions $\avNcoll$ is determined from a Monte Carlo Glauber model~\cite{Alver:2008aq,Loizides:2014vua,Loizides:2017ack} and related to the average nuclear overlap 
 $\avTaa = \avNcoll/\sigma^{\rm NN}_{\rm inel}$, where $\sigma^{\rm NN}_{\rm inel}$ is the total inelastic nucleon-nucleon cross section.
The yields measured in \AaAa\ collisions, as well as $\avNcoll$, depend on the collision centrality, and $\RAA$\com{ defined in \Eq{eq:raa}} is constructed to be unity in the absence of nuclear effects where particle production is dominated by hard processes. 
The collision centrality is expressed in percentiles of the total hadronic cross section, with the highest (lowest) centrality  0\% (100\%) refering to the most central (peripheral) collisions with zero (maximal) impact parameter.
Experimentally, centrality is typically determined by ordering events according to multiplicity or energy deposition  in a limited rapidity range and by fitting the corresponding distribution with a Glauber-based model of particle production~\cite{Abelev:2013qoq}. 

Numerous measurements of $\RAA$ reported by experiments at the Relativistic Heavy-Ion Collider (RHIC)~\cite{Adcox:2001jp,Adler:2002xw,Adcox:2002pe,Adler:2003qi,Adams:2003kv,Back:2003qr,Alver:2005nb,Adler:2006bw,Adare:2008cg,Adare:2012wg} and at the Large Hadron Collider (LHC)~\cite{Aamodt:2010jd,Abelev:2012hxa,CMS:2012aa,Aad:2015wga,Khachatryan:2016odn,raa5tevpaper} revealed that high-$\pT$ particle production is suppressed strongly in central collisions, and that the suppression reduces with decreasing centrality.
Furthermore, control measurements of possible nuclear modification arising from the initial state in \dAu\ and \pPb\ collisions~\cite{Arsene:2003yk,Adams:2003im,ALICE:2012mj,Abelev:2014dsa,Khachatryan:2015xaa,Aad:2016zif} and with electromagnetic probes in \AaAa\ collisions~\cite{Chatrchyan:2011ua,Afanasiev:2012dg,Chatrchyan:2012vq,Chatrchyan:2012nt,Aad:2015lcb}~(which should not be affected by partonic matter) demonstrated that the observed suppression is due to final state interactions, such as parton energy loss. 
Contrary to expectations, $\RAA$ was also found to be below unity at high $\pT$ in peripheral collisions\com{~(even when considering the relative large global uncertainties)}, reaching an approximately constant value of about $0.80$ above $3$ GeV/$c$ in $80$--$92$\% \AuAu\ collisions at $\snn=0.2$~TeV~\cite{Adare:2012wg} and about $0.75$ above $10$ GeV/$c$ in 70--90\% \PbPb\ collisions at $\snn=5.02$~TeV~\cite{Khachatryan:2016odn}.
In a final-state dominated scenario, such differences relative to unity imply a large jet quenching parameter for peripheral collisions, up to an order of magnitude larger than for cold nuclear matter~\cite{Dainese:2004te}, and consequently raise expectations of the relevance of parton energy loss even in small collision systems~\cite{Zhang:2013oca,Tywoniuk:2014hta,Shen:2016egw}. 
However, it has been pointed out recently~\cite{Morsch:2017brb} that event selection and geometry biases --- just like those discussed for \pPb\ collisions~\cite{Adam:2014qja} --- can cause an apparent suppression of \RAA\ in peripheral collisions, even in the absence of nuclear effects, while self-normalized coincidence observables~\cite{Adam:2015doa,Acharya:2017okq} are not affected.

The impact parameter of individual \NN\ collisions is correlated to the overall collision geometry leading to an \NN\ impact parameter bias in the transverse plane~\cite{Jia:2009mq}, for peripheral collisions the \NN\ impact parameter is biased towards larger values.
Centrality classification based on multiplicity can bias the mean multiplicity of individual nucleon--nucleon (\NN) collisions, and hence the yield of hard processes in \AaAa\ collisions due to correlated soft and hard particle production, amplifying the inherent \NN\ impact parameter bias.
The presence of the multiplicity bias in peripheral \PbPb\ collisions was already demonstrated in \Ref{Adam:2014qja} showing
the averaged multiplicity of the Glauber-NBD fit is lower than the average number of ancestors times the mean multiplicity of NBD (left panel of figure~8 in~\Ref{Adam:2014qja}).
In the present paper, we aim to study its relevance on charged-particle spectra in \PbPb\ collisions at $\snn=5.02$ TeV, in 20 centrality classes ranging from 0--5\% to 95--100\% collisions.
The spectra at midrapidity are measured in the range $0.15<\pT<30$~\gevc\ except for the 95--100\% class, where it is $0.15<\pT<20$~\gevc. 
Using the charged-particle spectra from \pp\ collisions at the same energy~\cite{raa5tevpaper}, we construct the nuclear-modification factor and study the centrality dependence of its average at high \pT\com{in $8<\pT<20$~\gevc}, as well as its slope at low and high \pT\com{within $8$--$20$~\gevc\ and within $0.5$--$2$~\gevc}.
To reduce uncertainties related to event selection and normalization, which are particularly large for peripheral collisions, we also provide the ratio of $\RAA$ in adjacent centrality intervals, defined as
\begin{equation}
\RPO \equiv \RPO^{i} = \frac{\RAA^{i}}{\RAA^{i+1}} \, = \frac{\avNcoll^{i+1}}{\avNcoll^i} \, \frac{{\rm d}N_{{\rm ch}}^{{\rm AA},i}/{\rm d}\pT}{{\rm d}N_{\rm ch}^{{\rm AA},i+1}/{\rm d}\pT}\,,
\label{eq:rpo}
\end{equation}
where $i+1$ denotes a 5\% more central centrality class than $i$. 
The definition of $\RPO$ corresponds approximately to the change of $\log \RAA$ with centrality, and its value would be constant for an exponential dependence.

Similar to $\RAA$, we quantify the centrality dependence of the average $\RPO$ at high \pT, as well as its slope at low and at high \pT.
Where possible, the results are compared to a PYTHIA-based model of independent pp collisions without nuclear modification~\cite{Morsch:2017brb}.
The remainder of the paper is structured as follows:
\Section{sec:setup} describes the experimental setup.
\Section{sec:ana} describes the charged particle measurement with emphasis on corrections and uncertainties related to the most peripheral collisions.
\Section{sec:results} describes the results.
\Section{sec:summary} provides a summary of our findings.

\section{Experimental setup}
\label{sec:setup}
The ALICE detector is described in detail in \Ref{Aamodt:2008zz}, and a summary of its performance can be found in \Ref{Abelev:2014ffa}.
Charged-particle reconstruction at midrapidity is based on tracking information from the Inner Tracking System (ITS) and the Time Projection Chamber (TPC), both located inside a solenoidal magnetic field of $0.5$~T parallel to the beam axis.

The ITS~\cite{Aamodt:2010aa} consists of three sub-detectors, each composed of two layers to measure the trajectories of charged particles and to reconstruct primary vertices. 
The two innermost layers are the Silicon Pixel Detectors (SPD), the middle two layers are Silicon Drift Detectors (SDD), the outer two layers are Silicon Strip Detectors (SSD).

The TPC~\cite{Alme:2010ke} is a large (90~${\rm m}^{3})$ cylindrical drift detector. 
It covers a pseudorapidity range of $|\eta|<0.9$ over full azimuth, providing up to 159 reconstructed space points per track.
Charged particles originating from the primary vertex can be reconstructed down to $\pT\approx 100$~MeV/$c$.
The relative \pT\ resolution depends on momentum, is approximately $4$\% at $0.15$~\gevc, $1$\% at $1$~\gevc\ and increases linearly approaching $4$\% at $50$~\gevc.

The \pp\ and \PbPb\ collision data at \snn\ = 5.02~TeV were recorded in 2015.
In total, about $110 \cdot 10^6$ pp and $ 25 \cdot 10^6$ \PbPb\ events satisfying the minimum bias trigger and a number of offline event selection criteria were used in the analysis.
The minimum-bias trigger required a signal in both, the V0-A and V0-C, scintillator arrays, covering $2.8<\eta<5.1$ and $-3.7<\eta<-1.7$, respectively~\cite{Abbas:2013taa}.
Beam background events were rejected efficiently by exploiting the timing signals in the V0 detectors, and in \PbPb\ collisions also by using the two Zero Degree Calorimeters~(ZDCs).
The latter are positioned close to beam rapidity on both sides of the interaction point.

\section{Data analysis}
\label{sec:ana}
The measurements of charged-particle spectra in \pp\ and \PbPb\ collisions at $\snn=5.02$~TeV are described in detail in Ref.~\cite{raa5tevpaper}.

The collision point or primary event vertex was determined from reconstructed tracks. 
If no vertex was found using tracks, the vertex reconstruction was performed using track segments constructed from the two innermost layers of the ITS. 
Events with a reconstructed vertex within $\pm$10~cm from the centre of the detector along the beam direction are used to ensure a uniform acceptance and reconstruction efficiency at midrapidity.

Primary charged particles~\cite{ALICE-PUBLIC-2017-005} were measured in the kinematic range of $|\eta|<0.8$ and $0.15 < \pt < 30$~\gevc.  
The detector simulations were performed using the PYTHIA~\cite{Sjostrand:2006za} and HIJING~\cite{Wang:1991hta} Monte Carlo event generators with GEANT3~\cite{geant3ref2} for modeling the detector response.
Track-level corrections include acceptance, efficiency, purity and \pT\ resolution, which were obtained using an improved method tuned on data to reduce the systematic uncertainties related to particle species dependence (see Ref.~\cite{raa5tevpaper} for details).
Events are classified in percentiles of the hadronic cross-section using the sum of the amplitudes of the V0-A and V0-C signals~(V0M estimator)~\cite{Abelev:2013qoq}. 
The absolute scale of the centrality is defined by the range of 0--90\% centrality in which a Glauber-based multiplicity model is fitted to the V0M distribution.
The lower centrality limit of 90\% of this range with its corresponding V0M signal is denoted the anchor point~(AP).
The multiplicity for each particle source is modeled with a negative binomial distribution, where the effective number of independent particle production sources is described by a linear combination of the number of participants~(\Npart) and collisions~(\Ncoll). 
The AP was shifted by $\pm0.5$\%, leading to a systematic uncertainty in the normalization of the spectra of up to 6.7\% for the 85--90\% centrality class. 
Unlike previous measurements in \PbPb\ collisions, the analysis was not limited to 0--90\% most central events, where effects of trigger inefficiency and contamination by electromagnetic processes are negligible, but also included the 90--100\% most peripheral collisions.
The V0M value corresponding to 95\% of the hadronic cross section was determined by selecting either 95\% of the events given by the Glauber-NBD parametrization, or the number of events in the 0--90\% centrality class multiplied by the factor 95/90, where the latter is used as a variation to assess the systematic uncertainty of the approach.
The difference on the measured yields between the two ways was assigned as additional systematic uncertainty. 
For the centrality class 90--95\% (95--100\%) the combined uncertainty amounts to a fully correlated part of 10.8\% (11.7\%) on the normalization of the spectra and a 2.9\% (4.6\%) residual effect on the shape.

\begin{figure}[t]
\centering
\includegraphics[width=0.45\columnwidth]{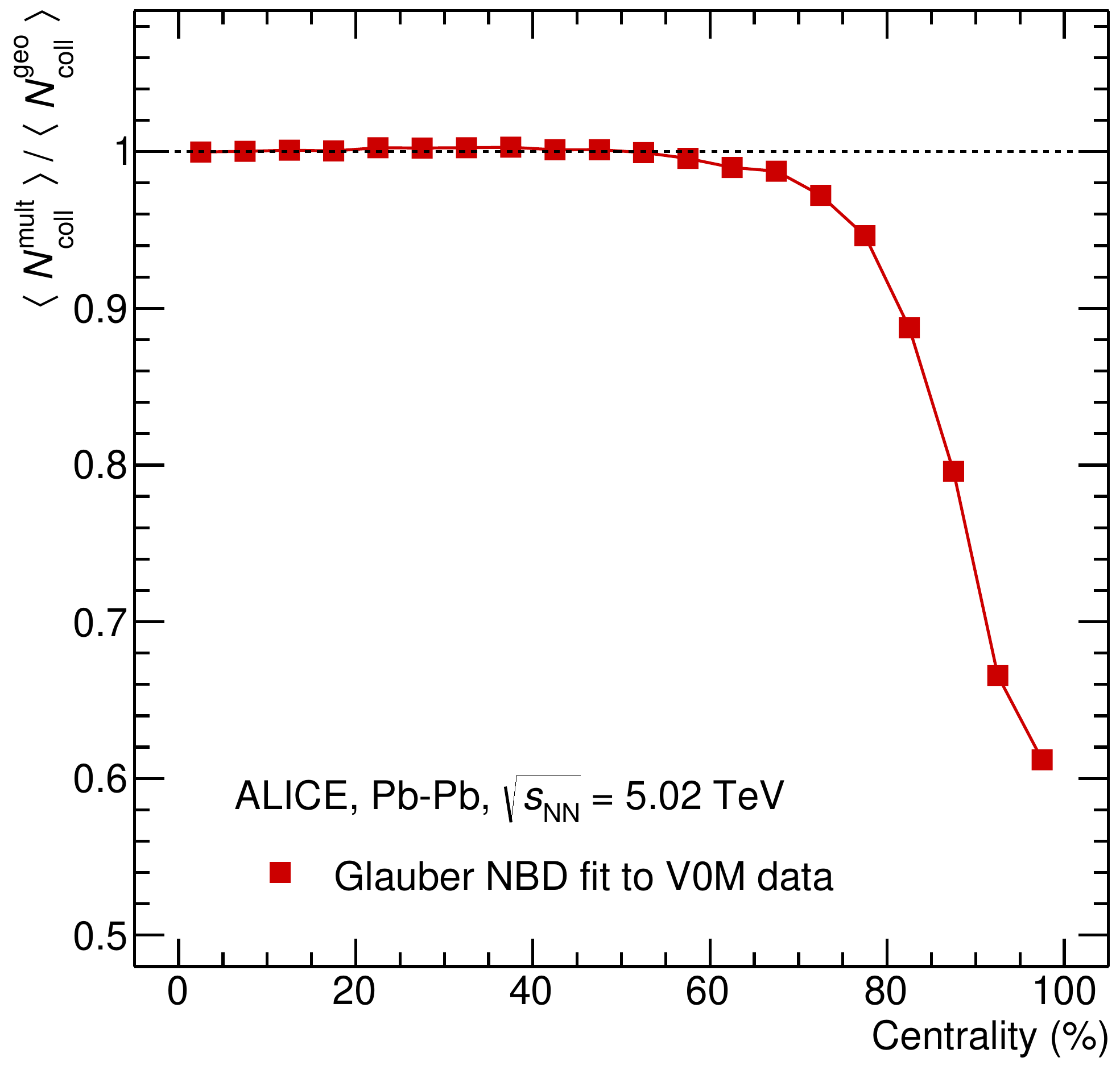} 
\caption{Ratio of number of collisions determined by slicing in multiplicity~($\Ncoll^{\rm mult}$) divided by the number of collisions determined directly from the impact parameter~($\Ncoll^{\rm geo}$).}
\label{fig:ncoll}
\end{figure}
\begin{table}[t!]
\centering
\begin{tabular}{@{}rrrr@{}} 
Centrality class \ & \avNpart & \avNcoll & \avTaa \ (mb$^{-1}$) \\ 
\hline
0--5\% 		&	$	382.3	\pm	2.4	$	&	$	1752	\pm	28	$	&	$	25.92	\pm	0.37	$	\\ 
5--10\%		&	$	329.1	\pm	5.0	$	&	$	1367	\pm	37	$	&	$	20.22	\pm	0.52	$	\\ 
10--15\%	&	$	281.1	\pm	5.2	$	&	$	1080	\pm	26	$	&	$	15.98	\pm	0.36	$	\\ 
15--20\%	&	$	239.4	\pm	5.2	$	&	$	 850	\pm	26	$	&	$	12.57	\pm	0.37	$	\\ 
20--25\%	&	$	202.7	\pm	4.6	$	&	$	 662	\pm	25	$	&	$	 9.79	\pm	0.36	$	\\ 
25--30\%	&	$	170.8	\pm	3.1	$	&	$	 513	\pm	16	$	&	$	 7.58	\pm	0.22	$	\\ 
30--35\%	&	$	142.5	\pm	3.0	$	&	$	 390	\pm	13	$	&	$	 5.77	\pm	0.18	$	\\ 
35--40\%	&	$	118.0	\pm	2.1	$	&	$	293.4	\pm	7.4	$	&	$	 4.34	\pm	0.11	$	\\ 
40--45\%	&	$	 96.3	\pm	2.0	$	&	$	215.2	\pm	6.4	$	&	$	 3.184	\pm	0.095	$	\\ 
45--50\%	&	$	 77.5	\pm	1.5	$	&	$	154.8	\pm	4.0	$	&	$	2.290	\pm	0.066	$	\\ 
50--55\%	&	$	 61.29	\pm	0.86	$	&	$	109.0	\pm	1.8	$	&	$	1.612	\pm	0.033	$	\\ 
55--60\%	&	$	 47.43	\pm	0.59	$	&	$	74.1	\pm	1.4	$	&	$	1.096	\pm	0.026	$	\\ 
60--65\%	&	$	 35.84	\pm	0.67	$	&	$	49.2	\pm	1.2	$	&	$	0.728	\pm	0.020	$	\\ 
65--70\%	&	$	26.19	\pm	0.56	$	&	$	31.6	\pm	1.1	$	&	$	0.468	\pm	0.018	$	\\ 
70--75\%	&	$	18.60	\pm	0.40	$	&	$	19.89	\pm	0.77	$	&	$	0.294	\pm	0.012	$	\\ 
75--80\%	&	$	12.78	\pm	0.32	$	&	$	12.19	\pm	0.46	$	&	$	0.1803	\pm	0.0075	$	\\ 
80--85\%	&	$	 8.50	\pm	0.23	$	&	$	 7.22	\pm	0.30	$	&	$	0.1068	\pm	0.0048	$	\\ 
85--90\%	&	$	 5.45	\pm	0.11	$	&	$	 4.12	\pm	0.13	$	&	$	0.0609	\pm	0.0021	$	\\ 
90--95\%	&	$	 3.31	\pm	0.19	$	&	$	 2.18	\pm	0.16	$	&	$	0.0323	\pm	0.0024	$	\\ 
95--100\%	&	$	 2.24	\pm	0.11	$	&	$	1.223	\pm	0.096	$	&	$	0.0181	\pm	0.0014	$	\\ 
\hline
\end{tabular}
\caption{\label{tab:1} Summary of the average \Npart, \Ncoll, \TAA\ for all centrality classes obtained by slicing the V0M amplitude distribution instead of the impact parameter. All uncertainties listed are systematic uncertainties. Statistical uncertainties are negligible.}
\end{table}

\begin{table}[t]
\centering
\begin{tabular}{@{}rrrr@{}} 
Centrality class & \dNchdeta  & $\mpt_{>0.15}$ \ (\gevc) & $\mpt_{>0}$ \ (\gevc) \\ 
\hline
0--5\%		&	$	1910	\pm	49	$	&	$	0.729	\pm	0.010	$	&	$	0.681	\pm	0.010	$	\\ 
5--10\%		&	$	1547	\pm	40	$	&	$	0.731	\pm	0.010	$	&	$	0.683	\pm	0.010	$	\\ 
10--15\%	&	$	1273	\pm	30	$	&	$	0.732	\pm	0.009	$	&	$	0.683	\pm	0.009	$	\\ 
15--20\%	&	$	1048	\pm	25	$	&	$	0.733	\pm	0.009	$	&	$	0.683	\pm	0.009	$	\\ 
20--25\%	&	$	863	\pm	19	$	&	$	0.730	\pm	0.009	$	&	$	0.678	\pm	0.008	$	\\ 
25--30\%	&	$	703	\pm	16	$	&	$	0.727	\pm	0.009	$	&	$	0.676	\pm	0.008	$	\\ 
30--35\%	&	$	568	\pm	13	$	&	$	0.723	\pm	0.008	$	&	$	0.671	\pm	0.008	$	\\ 
35--40\%	&	$	453	\pm	11	$	&	$	0.719	\pm	0.008	$	&	$	0.666	\pm	0.008	$	\\ 
40--45\%	&	$	356.6	\pm	8.4	$	&	$	0.710	\pm	0.008	$	&	$	0.657	\pm	0.008	$	\\ 
45--50\%	&	$	275.1	\pm	6.8	$	&	$	0.704	\pm	0.008	$	&	$	0.650	\pm	0.007	$	\\ 
50--55\%	&	$	208.5	\pm	5.6	$	&	$	0.695	\pm	0.008	$	&	$	0.640	\pm	0.008	$	\\ 
55--60\%	&	$	154.1	\pm	4.5	$	&	$	0.687	\pm	0.008	$	&	$	0.631	\pm	0.007	$	\\ 
60--65\%	&	$	111.4	\pm	3.5	$	&	$	0.676	\pm	0.007	$	&	$	0.619	\pm	0.007	$	\\ 
65--70\%	&	$	78.0	\pm	2.8	$	&	$	0.667	\pm	0.007	$	&	$	0.609	\pm	0.007	$	\\ 
70--75\%	&	$	53.1	\pm	2.1	$	&	$	0.659	\pm	0.007	$	&	$	0.599	\pm	0.007	$	\\ 
75--80\%	&	$	34.9	\pm	1.6	$	&	$	0.650	\pm	0.008	$	&	$	0.589	\pm	0.007	$	\\ 
80--85\%	&	$	22.0	\pm	1.4	$	&	$	0.636	\pm	0.014	$	&	$	0.575	\pm	0.013	$	\\ 
85--90\%	&	$	12.87	\pm	0.98	$	&	$	0.612	\pm	0.014	$	&	$	0.551	\pm	0.013	$	\\ 
90--95\%	&	$	6.46	\pm	0.78	$	&	$	0.574	\pm	0.017	$	&	$	0.516	\pm	0.015	$	\\ 
95--100\%	&	$	2.71	\pm	0.51	$	&	$	0.524	\pm	0.031	$	&	$	0.471	\pm	0.028	$	\\ 
\hline
\end{tabular}
\caption{\label{tab:2} Summary of the average \dNchdeta\ and \mpt\ in $|\eta| < 0.8 $ for all centrality classes. While $\mpt_{>0.15}$ is averaged over the measured range $0.15 < \pt < 10 $~GeV/$c$, $\mpt_{>0}$ is extrapolated to $\pt = 0$. All uncertainties listed are systematic uncertainties. Statistical uncertainties are negligible.}
\end{table}

The trigger and event-vertex reconstruction efficiency and the related systematic uncertainties for peripheral \PbPb\ collisions were estimated from simulations using HIJING and PYTHIA including single- and double-diffractive processes, but ignoring possible differences from nuclear effects. 
The V0M distribution in the simulations was reweighted with the measured V0M distribution.
The combined efficiency was found to be $0.985\pm0.015$ for the 90--95\% and $0.802\pm0.057$ for the 95--100\% centrality classes, respectively, while fully efficient for more central collisions. 
In addition, in the most peripheral bin a $\pt$-dependent signal loss of up to 14.7\% at low \pt\ is corrected for.
To account for diffractive processes in this correction and its systematic uncertainty, two limiting scenarios have been considered: 
a) the signal loss is assumed to be as in pp collisions in the V0M range of the 95--100\% bin;
b) only the fraction of events with a single nucleon--nucleon collision are corrected for assuming the signal loss from minimum-bias pp collisions.

Contamination of the peripheral bins by electromagnetic interactions was studied in the data by removing all events with small energy deposits in the neutron ZDCs.
The resulting change of the spectrum with the requirements of at least a five-neutron equivalent energy in both neutron ZDCs amounts to 5\% for the 95--100\% centrality class, 3\% for the 90-95\% class and 2\% for the 80-85\% and 85-90\% classes and is assigned as systematic uncertainty.
To account for contamination of the trigger from events without reconstructed vertex, those events are removed from the analysis and the effect is assigned as systematic uncertainty on the normalization~(6.8\% in the 95--100\% class and 0.5\% in the 90--95\% class).

Systematic uncertainties related to vertex selection, track selection, secondary-particle contamination, primary-particle composition, \pt\ resolution, material budget and tracking efficiency were estimated as described in Ref.~\cite{raa5tevpaper} and are assigned as bin-by-bin uncertainties.
The systematic uncertainties related to the centrality selection were estimated by a comparison of the \pT\ spectra when the limits of the centrality classes are shifted due to an uncertainty of $\pm$0.5\% in the fraction of the hadronic cross section used in the analysis.
They are split into two parts: one part that is fully correlated between the \pt\ bins assigned as a normalization uncertainty plus an additional part taking into account residual differences in the spectral shape assigned as a bin-by-bin uncertainty.
The overall normalization uncertainty of \RAA\ contains the uncertainty related to the centrality selection, the uncertainty of \Ncoll, the uncertainty of the trigger efficiency, the uncertainty of the trigger contamination and the normalization uncertainty of the pp reference spectrum added in quadrature.
Note that most uncertainties are correlated to a large extent between adjacent centrality bins leading to reduced uncertainties in \RPO.

Ordering events according to multiplicity introduces a bias relative to using the impact parameter in Glauber-based particle production models.
It is expected that part of the bias introduced by the ordering can be canceled in $\RAA$, when $\Ncoll$ is also obtained in the same way as in the data.
The difference relative to averaging over impact parameter is quantified in \Fig{fig:ncoll}, which shows the ratio of $\avNcoll$ by slicing either in multiplicity~(estimated using the V0M amplitude) $\avNcollmult$ or impact parameter $\avNcollgeo$, as carried out so far at the LHC.
The difference is below 5\% up to 80\% centrality, and then increases strongly up to 40\% for more peripheral classes.
The average quantities for a centrality class, such as the number of participants \Npart, the number of binary collisions \Ncoll\ and the nuclear overlap function \TAA, were obtained by averaging over the V0M multiplicity intervals, and are summarized in \Tab{tab:1}. 
For the calculation of $\RAA$ and $\RPO$ we use only those multiplicity averaged quantities.
As before~\cite{Abelev:2013qoq,Loizides:2017ack}, the uncertainties on the mean were obtained by changing the various ingredients of the Glauber MC model by one standard deviation.
The resulting relative uncertainties on the mean are below 6\%, however in particular for peripheral collisions the widths of the respective distributions are significantly larger. 

The charged particle multiplicity \dNchdeta\ and the average transverse momentum \mpt\ for all centrality intervals are listed in \Tab{tab:2}, values given for \dNchdeta and  $\mpt_{>0}$  are extrapolated to $\pt = 0$  using a modified Hagedorn function fitted to the data, as described in Ref. \cite{Acharya:2018eaq}.

\begin{figure}[t]
\centering
\includegraphics[width=0.95\textwidth]{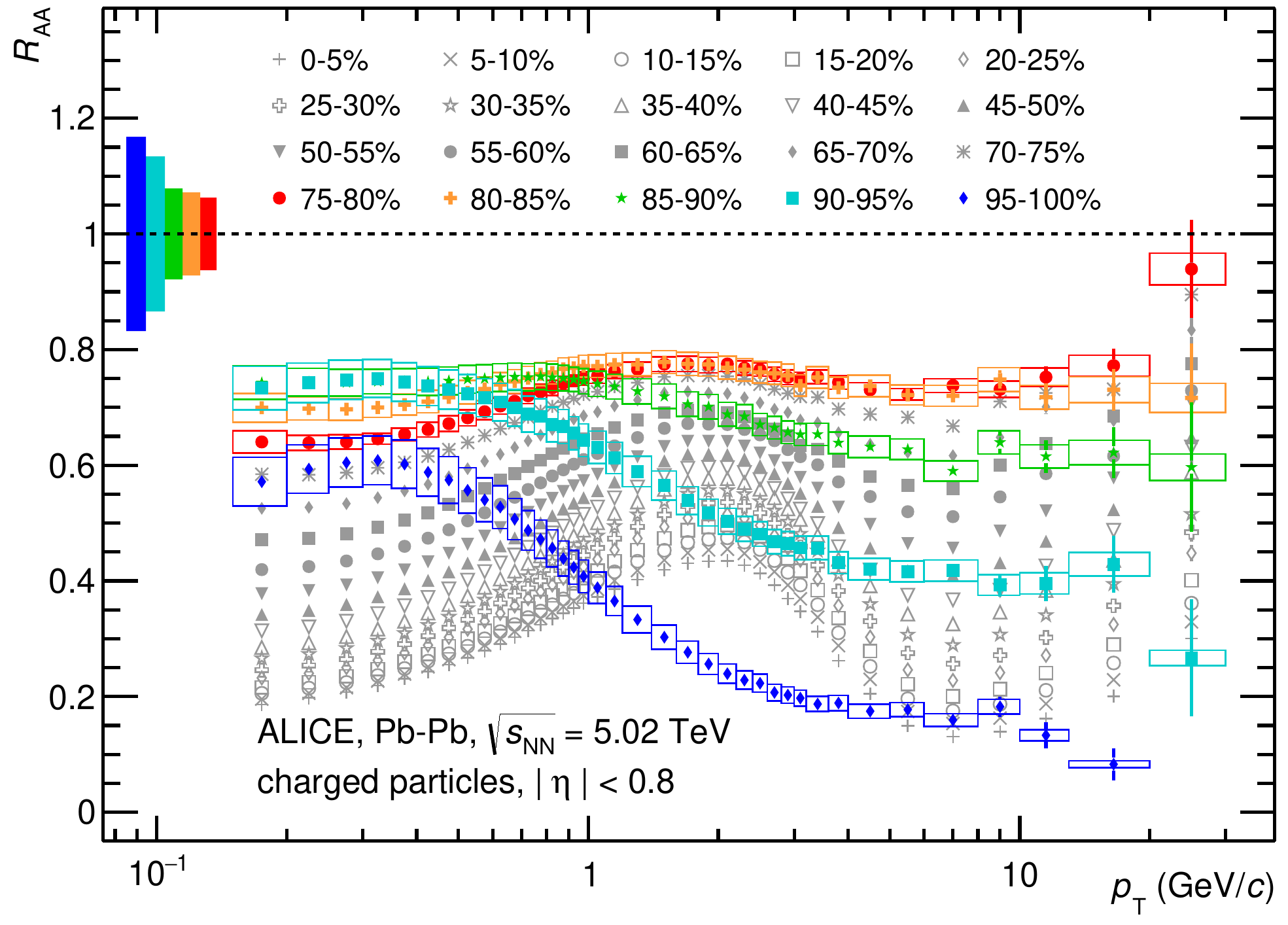} 
\caption{Nuclear-modification factor versus \pT\ for charged particles at midrapidity in \PbPb\ collisions at $\snn=5.02$ TeV for 5\%-wide centrality classes.
         The filled, coloured markers are for the five most peripheral classes, with the corresponding global uncertainties denoted close to $\pt = 0.1$~\gevc.
         Vertical error bars denote statistical uncertainties, while the boxes denote the systematic uncertainties. For visibility, the uncertainties are only drawn for the peripheral classes.}
\label{fig:raa}
\end{figure}

\section{Results}
\label{sec:results}
\Figure{fig:raa} presents the nuclear-modification factor, given in \Eq{eq:raa}, versus \pT\ for charged particles at midrapidity in \PbPb\ collisions at $\snn=5.02$ TeV for 5\%-wide centrality classes.
The focus of the presented analysis is mainly on the peripheral classes, which for convenience are displayed in filled, coloured symbols with their corresponding global uncertainties of about $10$--$20$\% denoted at $\pT{\sim}\,0.1$~\gevc.
As usual, if not otherwise stated, vertical error bars denote statistical uncertainties, while the boxes denote the systematic uncertainties.

From central to peripheral collisions $\RAA$ increases, which in particular above about $10$~\gevc\ can be understood as the progressive reduction of medium-induced parton energy loss.
Furthermore, the shape is similar from the most central up to the $80$--$85$\% centrality class, namely an increase at low \pt, a maximum around $2$--$3$~\gevc, related to radial flow, then a decrease with a local minimum at about $7$~\gevc, followed by a mild increase. 
Above $80$--$85$\% centrality, the evolution is different as already at low \pt\ the slope is negative and $\RAA$ decreases monotonously with increasing $\pT$. 
The change in behaviour seems to occur in the $75$--$85$\% interval, since the $80$--$85$\% $\RAA$ values appear to be the same or even lower than those of the $75$--$80$\% interval.  
For the most peripheral classes, the reduction of the nuclear modification factor with increasing $\pT$ is qualitatively similar to the one observed for low multiplicity \pPb~\cite{Adam:2014qja}\com{ and pp~\cite{ppspecpaaper}} collisions, indicating that the underlying bias towards more peripheral collisions with a reduced rate of hard scatterings per nucleon--nucleon collisions is the same.
If instead of using $\Ncoll^{\rm mult}$, we had used $\Ncoll^{\rm geo}$ in the normalization of $\RAA$, the results for peripheral collisions above 80\% would be even lower, namely by the ratio quantified in \Fig{fig:ncoll}. 

\begin{figure}[t]
\centering
 \begin{minipage}[t]{0.48\textwidth}
  \includegraphics[width=\textwidth]{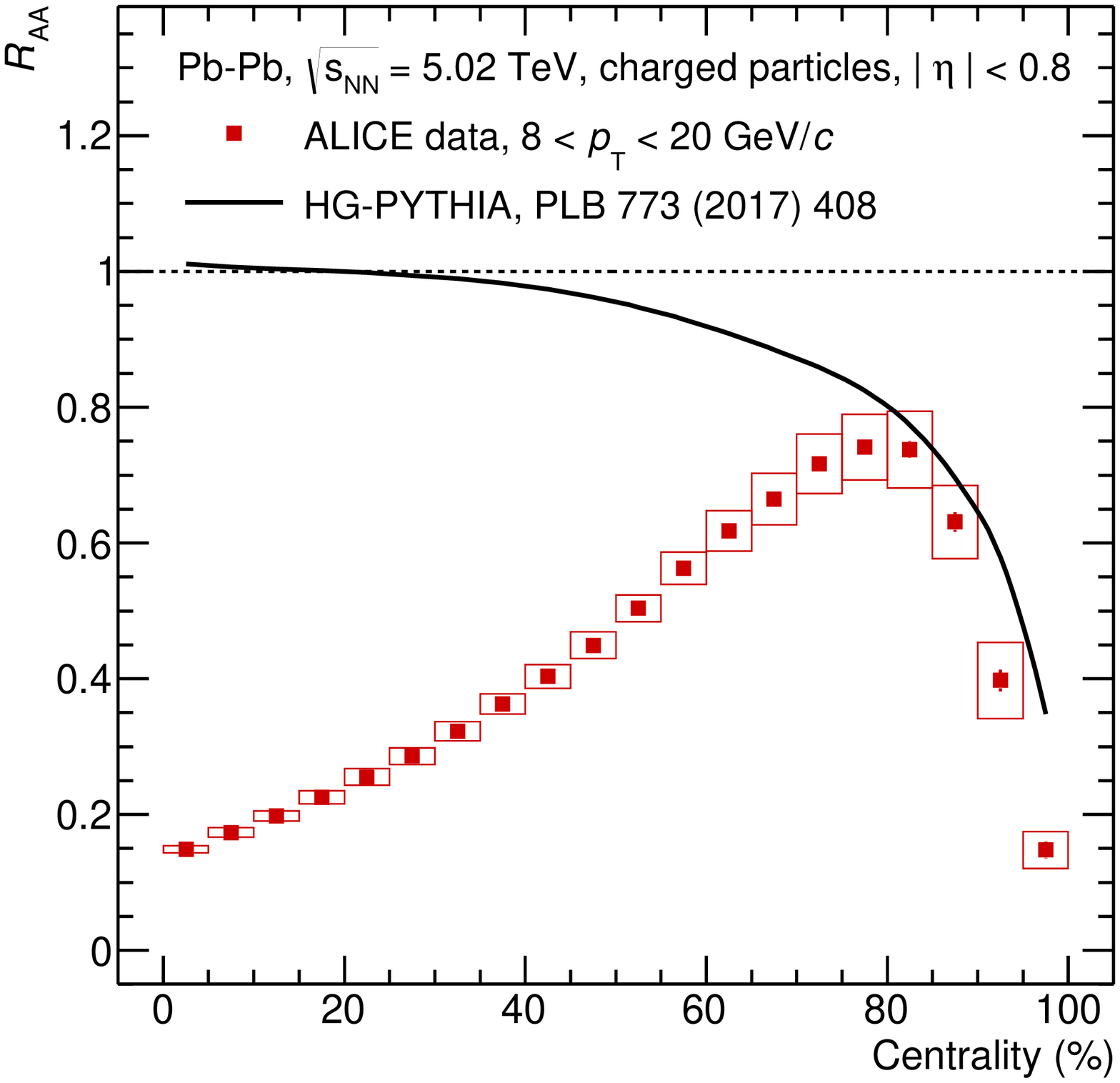} 
  \caption{Average $\RAA$ for $8<\pT<20$~\gevc\ versus centrality percentile in \PbPb\ collisions at $\snn=5.02$ TeV compared to predictions from HG-PYTHIA~\cite{Morsch:2017brb}.
           Vertical error bars denote statistical uncertainties, while the boxes denote the systematic uncertainties.}
  \label{fig:raafitconst}
 \end{minipage}
 \hspace{0.3cm}
 \begin{minipage}[t]{0.48\textwidth}
  \includegraphics[width=\textwidth]{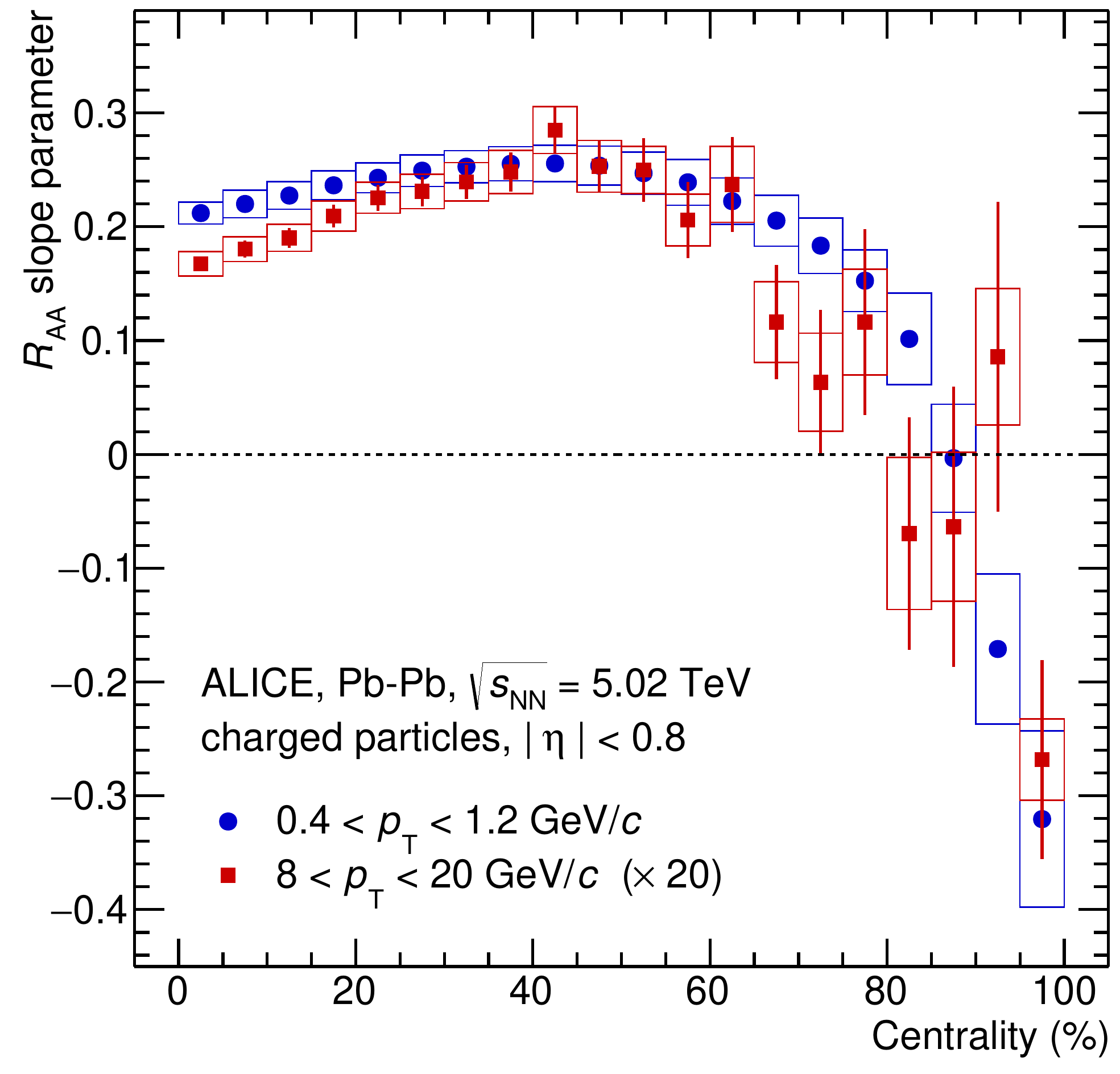} 
  \caption{Slope of $\RAA$ at low $\pT$~(in $0.4<\pT<1.2$~\gevc) and at high $\pT$~(in $8<\pT<20$~\gevc) scaled by factor 15 for visibility versus centrality percentile in \PbPb\ collisions at $\snn=5.02$ TeV. 
           Vertical error bars denote statistical uncertainties, while the boxes denote the systematic uncertainties.}
  \label{fig:raafitslope}
 \end{minipage}
\end{figure}

To quantify these observations we provide in \Fig{fig:raafitconst} the average $\RAA$ at high $\pt$\ (within $8<\pT<20$~\gevc), which increases smoothly from most central up to $70$--$75$\% centrality and drops strongly beyond the $80$--$85$\% centrality class.
The data are compared to the high $\pt$\ limit of a PYTHIA-based model~(HG-PYTHIA)~\cite{Morsch:2017brb}, which for every binary nucleon--nucleon collision superimposes a number of PYTHIA events incoherently without nuclear modification.
The essential feature of the model is that particle production per nucleon--nucleon collision originates from a fluctuating number of multiple partonic interactions depending on the nucleon--nucleon impact parameter.
Despite the fact that HG-PYTHIA is a rather simple approach, for $75$--$80$\% and more peripheral collisions, it describes the average $\RAA$ relatively well suggesting that the apparent suppression for peripheral collisions is not caused by parton energy loss, but rather by the event selection criteria imposed to determine the centrality of the collisions.
The data are significantly lower than the model calculation for the most peripheral centrality classes, possibly due to a significant contribution of diffraction, which is not modeled in HG-PYTHIA.
The slope of a linear fit to \RAA\ performed for $8<\pT<20$~\gevc, the region where the $\RAA$ in central collisions rises after its minimum, is
shown in \Fig{fig:raafitslope} as a function of centrality.
This high-\pT\ slope is positive and initially increasing mildly before decreasing with decreasing centrality up to about $80$\% centrality, beyond which it is close to zero, and then even is negative in the highest centrality class.
At low to intermediate $\pT$~(within $0.4$--$1.2$~\gevc), the regime which is strongly influenced by the hydrodynamic expansion, the \RAA\ exhibits another rise. 
The slope extracted in the $\pT$\ range $0.4$--$1.2$~\gevc\  is also shown in \Fig{fig:raafitslope}. The $\RAA$ at low and high \pT\ is consistent with being linearly dependent on \pT\ in the chosen fit ranges, resulting $\chi^2/\mbox{NDF}$ are below unity.
While the absolute values of the slopes are very different (note the normalisation), the shape of the centrality dependence of the slope at low $\pT$\ is remarkably similar to that extracted at high $\pT$.
This hints at a close correlation between these two regimes, possibly induced by the geometry or density dependence of parton energy loss on the one hand and collective expansion on the other hand.
In peripheral collisions, in particular above 90\% centrality, the low $\pT$ slope is negative, indicating that the very peripheral events are increasingly softer.

\begin{figure}[t]
\centering
\includegraphics[width=0.95\columnwidth]{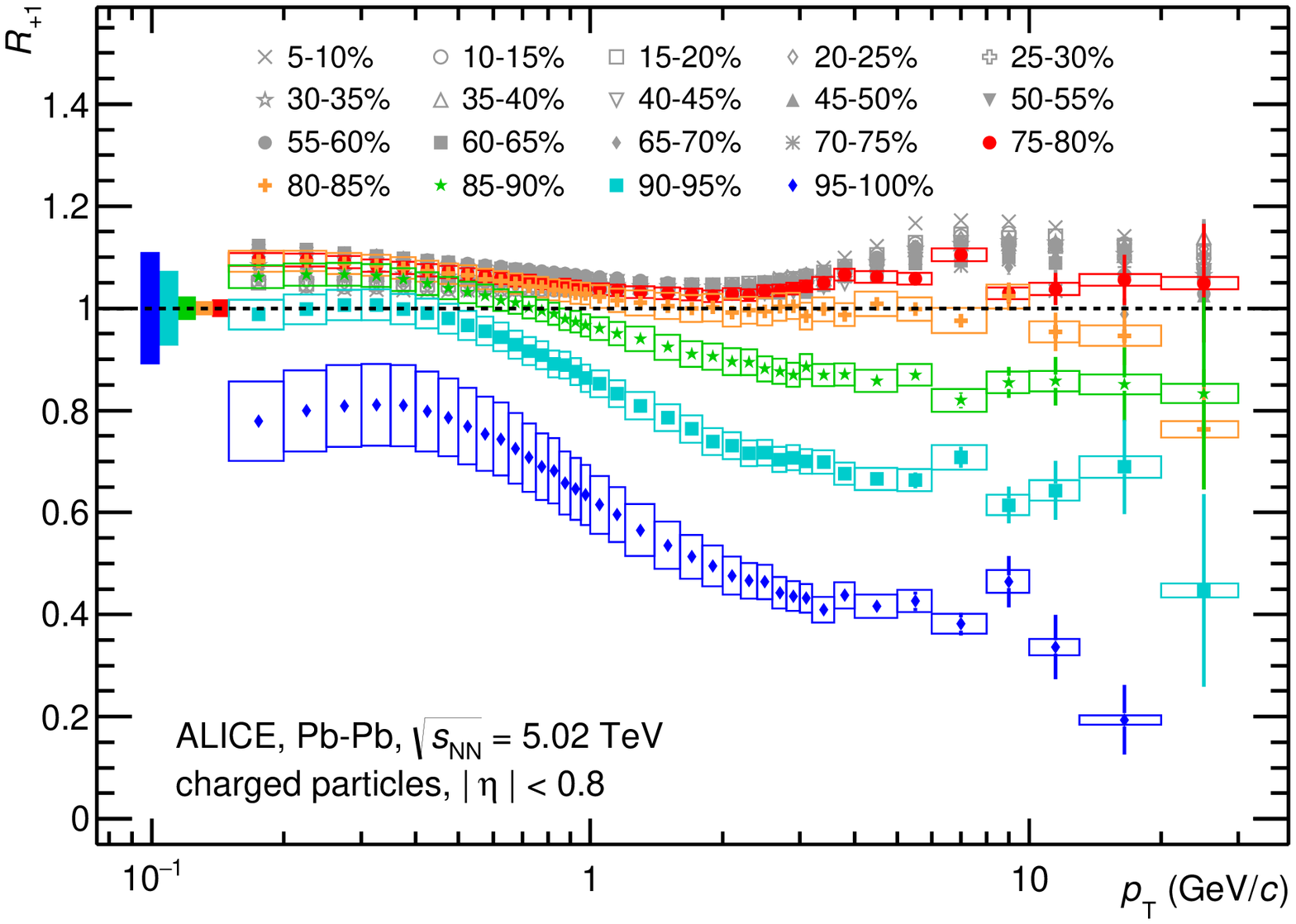} 
\caption{$\RPO$\ versus \pT\ for charged particles at midrapidity in \PbPb\ collisions at $\snn=5.02$ TeV. $\RPO$\ is defined as the ratio of $\Ncoll$ normalized spectra for a given centrality class relative to the 5\% more central class, see \Eq{eq:rpo}.
         The filled, coloured markers are for the 5 most peripheral classes, with the corresponding global uncertainties denoted close to $\pt = 0.1$~\gevc\ on the \pt-axis.
         Vertical error bars denote statistical uncertainties, while the boxes denote the systematic uncertainties. For visibility, the uncertainties are only drawn for the peripheral classes.}
\label{fig:rpo}
\end{figure}

\begin{figure}[t]
\centering
 \begin{minipage}[t]{0.48\textwidth}
  \includegraphics[width=\textwidth]{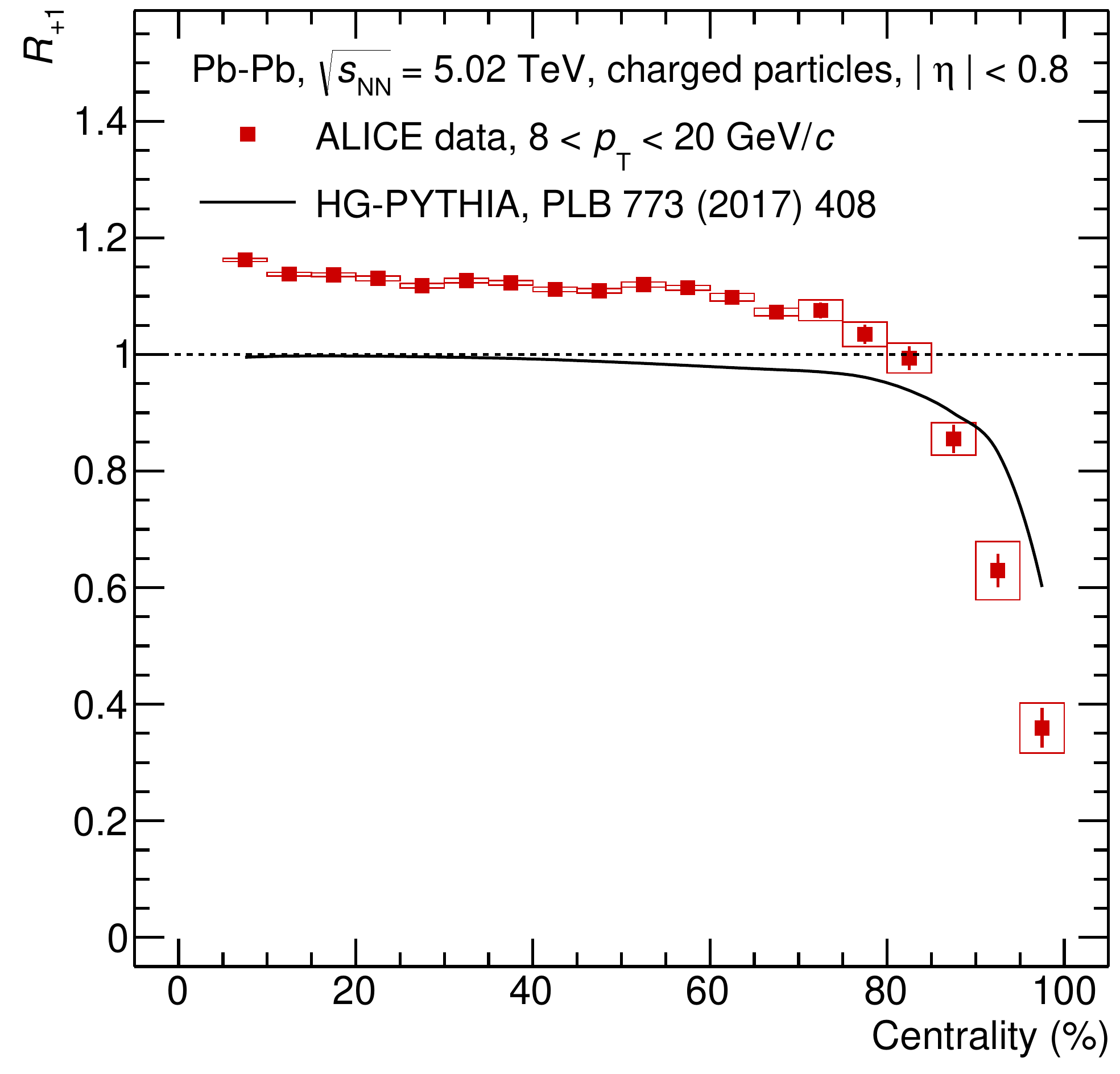} 
  \caption{Average $\RPO$ for $8<\pT<20$~\gevc\ versus centrality percentile in \PbPb\ collisions at $\snn=5.02$ TeV compared to predictions from HG-PYTHIA~\cite{Morsch:2017brb}.
           Vertical error bars denote statistical uncertainties, while the boxes denote the systematic uncertainties.}
  \label{fig:rpcfitconst}
 \end{minipage}
 \hspace{0.3cm}
 \begin{minipage}[t]{0.48\textwidth}
  \includegraphics[width=\textwidth]{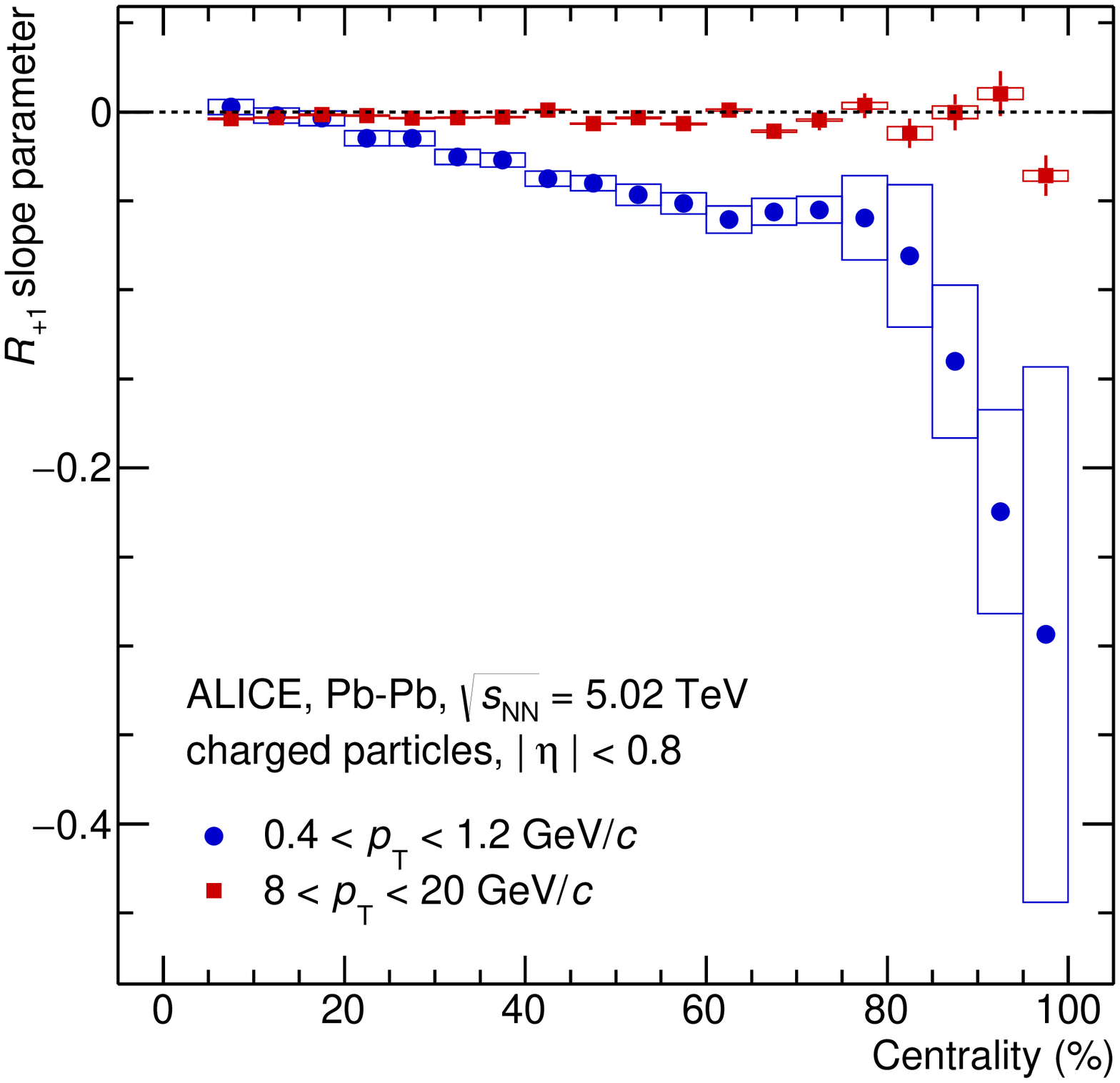} 
  \caption{Slope of $\RPO$ at low $\pT$ (in $0.4<\pT<1.2$~\gevc) and at high $\pT$ (in $8<\pT<20$~\gevc) versus centrality percentile in \PbPb\ collisions at $\snn=5.02$ TeV.
           Vertical error bars denote statistical uncertainties, while the boxes denote the systematic uncertainties.}
  \label{fig:rpcfitslope}
 \end{minipage}
\end{figure}

In order to study the shape evolution of $\RAA$ in more detail, we compute the ratio of adjacent centrality intervals, as given by \Eq{eq:rpo}.
In this way a large part of the global uncertainties as well as of the systematic uncertainties cancel. 
\Figure{fig:rpo} presents $\RPO$ versus \pT\ for charged particles at midrapidity in \PbPb\ collisions at $\snn=5.02$ TeV for 5\%-wide centrality classes.
As for $\RAA$ the peripheral collisions are displayed in colour, with their corresponding global uncertainties, which are significantly smaller than for $\RAA$ except for the most peripheral class, denoted around 0.1 on the abscissa.
The ratio is found to be nearly identical for 0--5\% central to 70--75\% peripheral collisions (14 curves) within $10$\%. 
In addition, in this centrality range, the ratio is only slightly $\pt$-dependent, although explained typically by distinct mechanism~(radial flow at low $\pt$ and energy loss at high $\pt$). 
For more peripheral collisions, however, the $\RPO$ changes significantly and reduces to about $0.4$ for most peripheral collisions. 
While the quenching power of the medium apparently only gradually changes for about 75\% of the \PbPb\ cross-section, the sudden drop for more than 75\% peripheral collisions can hardly be explained by an increase in quenching.

The evolution of the $\RPO$ at high $\pT$ with centrality is characterized by taking the average $\RPO$ for $8<\pt<20$~\gevc, shown in \Fig{fig:rpcfitconst}.
The average is about $1.14$, slightly decreasing with decreasing centrality and beyond 75\% centrality falls strongly, similar to predictions from HG-PYTHIA. 
An approximate constant value for $\RPO$ up to about 60\% centrality implies an exponential dependence on centrality.

\Figure{fig:rpcfitslope} shows the slope of a linear fit to the low momentum region (0.4--1.2~\gevc) and the high-momentum region~($8<\pt<20$~\gevc) of $\RPO$.
In the chosen fit ranges, the $\RPO$ can be fitted by a linear function with $\chi^2/\mbox{NDF} < 1$.
At low momentum, the slope of $\RPO$ exhibits a mild centrality dependence, related to the reduced strength of radial flow, dropping strongly for peripheral collisions above 80\%, as expected from ordering events according to multiplicity.
At high momentum, the slope is non-zero, $-0.0031 \pm 0.0006$, and within the uncertainties not dependent on centrality.

\section{Summary}
\label{sec:summary}
Charged-particle spectra at midrapidity were measured in Pb--Pb collisions at a centre-of-mass energy per nucleon pair of $\snn = 5.02$ TeV and presented in centrality classes ranging from the most central~(0--5\%) to the most peripheral~(95--100\%) collisions.
Measurements beyond the 90\% peripheral collisions at the LHC are presented for the first time.
For a consistent treatment of the most peripheral collisions the number of binary collisions was calculated from a Glauber model in intervals of multiplicity rather than in impact parameter~(\Fig{fig:ncoll}).
Possible medium effects were quantified by comparing the measured spectra with those from proton--proton collisions normalized by the number of independent nucleon--nucleon collisions obtained from a Glauber model~(\Fig{fig:raa}).
At large transverse momenta~($8<\pT<20$~\gevc), the average $\RAA$ increases from about $0.15$ in the 0--5\% most central collisions to a maximum value of about $0.8$ in the 75--85\% peripheral collisions, beyond which it strongly falls off to below $0.2$ for the most peripheral collisions~(\Fig{fig:raafitconst}).
Furthermore, \RAA\ initially exhibits a positive slope as a function of $\pT$ in the $8$--$20$~\gevc\ interval, while for collisions beyond the 80\% class the slope is negative (\Fig{fig:raafitslope}).
The shape of the slope extracted at low $\pT$, within $0.4$--$1.2$~\gevc, is remarkably similar, indicating that there may be a close correlation between these two regimes. 
To reduce uncertainties related to event selection and normalization, the ratio of $\RAA$ in adjacent centrality intervals was measured~(\Fig{fig:rpo}).
Up to about 60\% peripheral collisions, this ratio is fairly constant, even as a function of $\pT$.
It then starts to decrease and finally, for centralities beyond 75\%, it falls off strongly~(\Fig{fig:rpcfitconst}) with its slopes at low and high momentum varying only mildly or not at all except for the most peripheral centrality intervals~(\Fig{fig:rpcfitslope}).

The trends observed in peripheral collisions are consistent with a simple PYTHIA-based model without nuclear modification, demonstrating that biases caused by the event selection and collision geometry can lead to an apparent suppression in peripheral collisions.
This explains the contradictory and hard to reconcile observation that \RAA\ is below unity in peripheral \PbPb, but equal to unity in minimum-bias \pPb\ collisions despite similar charged-particle multiplicities.
With a correct treatment of the biases a smooth transition between \PbPb\ and minimum-bias \pPb\ collisions is expected without the need to involve parton energy loss in peripheral collisions. 
Without such treatment, the measurement and interpretation of $\RAA$ in peripheral collisions, in particular above 80\% centrality, have complications similar to \pPb\ collisions, where the observable was named $Q_{\rm pPb}$~\cite{Adam:2014qja} to distinguish it from the unbiased nuclear modification factor.

\newenvironment{acknowledgement}{\relax}{\relax}
\begin{acknowledgement}
\section*{Acknowledgements}
\input{fa_2018-05-11.tex}
\end{acknowledgement}

\bibliographystyle{utphys}
\bibliography{biblio}{}
\newpage
\appendix
\section{The ALICE Collaboration}
\label{app:collab}
\input{Alice_Authorlist_2018-May-04.tex}
\end{document}

%% file: commands.tex
\newcommand{\RPO}          {\ensuremath{R_{\rm +1}}}

\newcommand{\RAA}          {\ensuremath{R_{\rm AA}}}
\newcommand{\TAA}          {\ensuremath{T_{\rm AA}}}

\newcommand{\Ncoll}        {\ensuremath{N_{\rm coll}}}
\newcommand{\Ncollmult}    {\ensuremath{N_{\rm coll}^{\rm mult}}}
\newcommand{\Ncollgeo}    {\ensuremath{N_{\rm coll}^{\rm geo}}}
\newcommand{\Npart}        {\ensuremath{N_{\rm part}}}

\newcommand{\pp}           {pp}

\newcommand{\NN}           {\mbox{NN}}
\newcommand{\AaAa}         {\mbox{AA}}
\newcommand{\PbPb}         {\mbox{Pb--Pb}}
\newcommand{\AuAu}         {\mbox{Au--Au}}

\newcommand{\pPb}          {\mbox{p--Pb}}

\newcommand{\dAu}          {\mbox{d--Au}}

\newcommand{\dNchdeta}     {\ensuremath{\mathrm{d}N_\text{ch}/\mathrm{d}\eta}}

\newcommand{\gevc}         {\ensuremath{\mathrm{GeV}/c}}

\newcommand{\pt}           {\ensuremath{p_{\mathrm{T}}}}
\newcommand{\pT}           {\ensuremath{p_{\mathrm{T}}}}
\newcommand{\mpt}          {\ensuremath{\langle\pt\rangle}}

\newcommand{\snn}          {\ensuremath{\sqrt{s_{\mathrm{NN}}}}}

\newcommand{\avg}[1]       {\ensuremath{\left\langle#1\right\rangle}}

\newcommand{\avNcoll}      {\ensuremath{\avg{\Ncoll}}}
\newcommand{\avNcollmult}  {\ensuremath{\avg{\Ncollmult}}}
\newcommand{\avNcollgeo}   {\ensuremath{\avg{\Ncollgeo}}}
\newcommand{\avNpart}      {\ensuremath{\avg{\Npart}}}

\newcommand{\avTaa}        {\ensuremath{\avg{\TAA}}}

\newcommand{\Fig}[1]       {Fig.~\ref{#1}}
\newcommand{\Figure}[1]    {Figure~\ref{#1}}

\newcommand{\Section}[1]   {Section~\ref{#1}}

\newcommand{\Eq}[1]        {Eq.~\ref{#1}}

\newcommand{\Tab}[1]       {Tab.~\ref{#1}}

\newcommand{\Ref}[1]       {Ref.~\cite{#1}}

\newcommand{\com}[1]       {}
\iflatexdiff

\renewcommand{\xout}[1]    {\textcolor{red}{\sout{#1}}}
 
\newcommand{\old}[1]       {{\textcolor{red}{\sout{#1}}}}

\else

\renewcommand{\xout}[1]    {}
\newcommand{\old}[1]       {\relax}

\fi

%% file: fa_2018-05-11.tex

The ALICE Collaboration would like to thank all its engineers and technicians for their invaluable contributions to the construction of the experiment and the CERN accelerator teams for the outstanding performance of the LHC complex.
The ALICE Collaboration gratefully acknowledges the resources and support provided by all Grid centres and the Worldwide LHC Computing Grid (WLCG) collaboration.
The ALICE Collaboration acknowledges the following funding agencies for their support in building and running the ALICE detector:
A. I. Alikhanyan National Science Laboratory (Yerevan Physics Institute) Foundation (ANSL), State Committee of Science and World Federation of Scientists (WFS), Armenia;
Austrian Academy of Sciences and Nationalstiftung f\"{u}r Forschung, Technologie und Entwicklung, Austria;
Ministry of Communications and High Technologies, National Nuclear Research Center, Azerbaijan;
Conselho Nacional de Desenvolvimento Cient\'{\i}fico e Tecnol\'{o}gico (CNPq), Universidade Federal do Rio Grande do Sul (UFRGS), Financiadora de Estudos e Projetos (Finep) and Funda\c{c}\~{a}o de Amparo \`{a} Pesquisa do Estado de S\~{a}o Paulo (FAPESP), Brazil;
Ministry of Science \& Technology of China (MSTC), National Natural Science Foundation of China (NSFC) and Ministry of Education of China (MOEC) , China;
Ministry of Science and Education, Croatia;
Ministry of Education, Youth and Sports of the Czech Republic, Czech Republic;
The Danish Council for Independent Research | Natural Sciences, the Carlsberg Foundation and Danish National Research Foundation (DNRF), Denmark;
Helsinki Institute of Physics (HIP), Finland;
Commissariat \`{a} l'Energie Atomique (CEA) and Institut National de Physique Nucl\'{e}aire et de Physique des Particules (IN2P3) and Centre National de la Recherche Scientifique (CNRS), France;
Bundesministerium f\"{u}r Bildung, Wissenschaft, Forschung und Technologie (BMBF) and GSI Helmholtzzentrum f\"{u}r Schwerionenforschung GmbH, Germany;
General Secretariat for Research and Technology, Ministry of Education, Research and Religions, Greece;
National Research, Development and Innovation Office, Hungary;
Department of Atomic Energy Government of India (DAE), Department of Science and Technology, Government of India (DST), University Grants Commission, Government of India (UGC) and Council of Scientific and Industrial Research (CSIR), India;
Indonesian Institute of Science, Indonesia;
Centro Fermi - Museo Storico della Fisica e Centro Studi e Ricerche Enrico Fermi and Istituto Nazionale di Fisica Nucleare (INFN), Italy;
Institute for Innovative Science and Technology , Nagasaki Institute of Applied Science (IIST), Japan Society for the Promotion of Science (JSPS) KAKENHI and Japanese Ministry of Education, Culture, Sports, Science and Technology (MEXT), Japan;
Consejo Nacional de Ciencia (CONACYT) y Tecnolog\'{i}a, through Fondo de Cooperaci\'{o}n Internacional en Ciencia y Tecnolog\'{i}a (FONCICYT) and Direcci\'{o}n General de Asuntos del Personal Academico (DGAPA), Mexico;
Nederlandse Organisatie voor Wetenschappelijk Onderzoek (NWO), Netherlands;
The Research Council of Norway, Norway;
Commission on Science and Technology for Sustainable Development in the South (COMSATS), Pakistan;
Pontificia Universidad Cat\'{o}lica del Per\'{u}, Peru;
Ministry of Science and Higher Education and National Science Centre, Poland;
Korea Institute of Science and Technology Information and National Research Foundation of Korea (NRF), Republic of Korea;
Ministry of Education and Scientific Research, Institute of Atomic Physics and Romanian National Agency for Science, Technology and Innovation, Romania;
Joint Institute for Nuclear Research (JINR), Ministry of Education and Science of the Russian Federation and National Research Centre Kurchatov Institute, Russia;
Ministry of Education, Science, Research and Sport of the Slovak Republic, Slovakia;
National Research Foundation of South Africa, South Africa;
Centro de Aplicaciones Tecnol\'{o}gicas y Desarrollo Nuclear (CEADEN), Cubaenerg\'{\i}a, Cuba and Centro de Investigaciones Energ\'{e}ticas, Medioambientales y Tecnol\'{o}gicas (CIEMAT), Spain;
Swedish Research Council (VR) and Knut \& Alice Wallenberg Foundation (KAW), Sweden;
European Organization for Nuclear Research, Switzerland;
National Science and Technology Development Agency (NSDTA), Suranaree University of Technology (SUT) and Office of the Higher Education Commission under NRU project of Thailand, Thailand;
Turkish Atomic Energy Agency (TAEK), Turkey;
National Academy of  Sciences of Ukraine, Ukraine;
Science and Technology Facilities Council (STFC), United Kingdom;
National Science Foundation of the United States of America (NSF) and United States Department of Energy, Office of Nuclear Physics (DOE NP), United States of America.

%% file: Alice_Authorlist_2018-May-04.tex

\begingroup
\small
\begin{flushleft}
S.~Acharya\Irefn{org139}\And 
F.T.-.~Acosta\Irefn{org20}\And 
D.~Adamov\'{a}\Irefn{org93}\And 
J.~Adolfsson\Irefn{org80}\And 
M.M.~Aggarwal\Irefn{org98}\And 
G.~Aglieri Rinella\Irefn{org34}\And 
M.~Agnello\Irefn{org31}\And 
N.~Agrawal\Irefn{org48}\And 
Z.~Ahammed\Irefn{org139}\And 
S.U.~Ahn\Irefn{org76}\And 
S.~Aiola\Irefn{org144}
% ...
\And
\renewcommand\labelenumi{\textsuperscript{\theenumi}~}

\section*{Affiliation notes}
\renewcommand\theenumi{\roman{enumi}}
\begin{Authlist}
\item \Adef{org*}Deceased
\item \Adef{orgI}Dipartimento DET del Politecnico di Torino, Turin, Italy
\item \Adef{orgII}M.V. Lomonosov Moscow State University, D.V. Skobeltsyn Institute of Nuclear, Physics, Moscow, Russia
\item \Adef{orgIII}Department of Applied Physics, Aligarh Muslim University, Aligarh, India
\item \Adef{orgIV}Institute of Theoretical Physics, University of Wroclaw, Poland
\end{Authlist}

\section*{Collaboration Institutes}
\renewcommand\theenumi{\arabic{enumi}~}
\begin{Authlist}
\item \Idef{org1}A.I. Alikhanyan National Science Laboratory (Yerevan Physics Institute) Foundation, Yerevan, Armenia
\item \Idef{org2}Bogolyubov Institute for Theoretical Physics, National Academy of Sciences of Ukraine, Kiev, Ukraine
\item \Idef{org3}Bose Institute, Department of Physics  and Centre for Astroparticle Physics and Space Science (CAPSS), Kolkata, India
\item \Idef{org4}Budker Institute for Nuclear Physics, Novosibirsk, Russia
\item \Idef{org5}California Polytechnic State University, San Luis Obispo, California, United States
\item \Idef{org6}Central China Normal University, Wuhan, China
\item \Idef{org7}Centre de Calcul de l'IN2P3, Villeurbanne, Lyon, France
\item \Idef{org8}Centro de Aplicaciones Tecnol\'{o}gicas y Desarrollo Nuclear (CEADEN), Havana, Cuba
\item \Idef{org9}Centro de Investigaci\'{o}n y de Estudios Avanzados (CINVESTAV), Mexico City and M\'{e}rida, Mexico
\item \Idef{org10}Centro Fermi - Museo Storico della Fisica e Centro Studi e Ricerche ``Enrico Fermi', Rome, Italy
\item \Idef{org11}Chicago State University, Chicago, Illinois, United States
\item \Idef{org12}China Institute of Atomic Energy, Beijing, China
\item \Idef{org13}Chonbuk National University, Jeonju, Republic of Korea
\item \Idef{org14}Comenius University Bratislava, Faculty of Mathematics, Physics and Informatics, Bratislava, Slovakia
\item \Idef{org15}COMSATS Institute of Information Technology (CIIT), Islamabad, Pakistan
\item \Idef{org16}Creighton University, Omaha, Nebraska, United States
\item \Idef{org17}Department of Physics, Aligarh Muslim University, Aligarh, India
\item \Idef{org18}Department of Physics, Pusan National University, Pusan, Republic of Korea
\item \Idef{org19}Department of Physics, Sejong University, Seoul, Republic of Korea
\item \Idef{org20}Department of Physics, University of California, Berkeley, California, United States
\item \Idef{org21}Department of Physics, University of Oslo, Oslo, Norway
\item \Idef{org22}Department of Physics and Technology, University of Bergen, Bergen, Norway
\item \Idef{org23}Dipartimento di Fisica dell'Universit\`{a} 'La Sapienza' and Sezione INFN, Rome, Italy
\item \Idef{org24}Dipartimento di Fisica dell'Universit\`{a} and Sezione INFN, Cagliari, Italy
\item \Idef{org25}Dipartimento di Fisica dell'Universit\`{a} and Sezione INFN, Trieste, Italy
\item \Idef{org26}Dipartimento di Fisica dell'Universit\`{a} and Sezione INFN, Turin, Italy
\item \Idef{org27}Dipartimento di Fisica e Astronomia dell'Universit\`{a} and Sezione INFN, Bologna, Italy
\item \Idef{org28}Dipartimento di Fisica e Astronomia dell'Universit\`{a} and Sezione INFN, Catania, Italy
\item \Idef{org29}Dipartimento di Fisica e Astronomia dell'Universit\`{a} and Sezione INFN, Padova, Italy
\item \Idef{org30}Dipartimento di Fisica `E.R.~Caianiello' dell'Universit\`{a} and Gruppo Collegato INFN, Salerno, Italy
\item \Idef{org31}Dipartimento DISAT del Politecnico and Sezione INFN, Turin, Italy
\item \Idef{org32}Dipartimento di Scienze e Innovazione Tecnologica dell'Universit\`{a} del Piemonte Orientale and INFN Sezione di Torino, Alessandria, Italy
\item \Idef{org33}Dipartimento Interateneo di Fisica `M.~Merlin' and Sezione INFN, Bari, Italy
\item \Idef{org34}European Organization for Nuclear Research (CERN), Geneva, Switzerland
\item \Idef{org35}Faculty of Electrical Engineering, Mechanical Engineering and Naval Architecture, University of Split, Split, Croatia
\item \Idef{org36}Faculty of Engineering and Science, Western Norway University of Applied Sciences, Bergen, Norway
\item \Idef{org37}Faculty of Nuclear Sciences and Physical Engineering, Czech Technical University in Prague, Prague, Czech Republic
\item \Idef{org38}Faculty of Science, P.J.~\v{S}af\'{a}rik University, Ko\v{s}ice, Slovakia
\item \Idef{org39}Frankfurt Institute for Advanced Studies, Johann Wolfgang Goethe-Universit\"{a}t Frankfurt, Frankfurt, Germany
\item \Idef{org40}Gangneung-Wonju National University, Gangneung, Republic of Korea
\item \Idef{org41}Gauhati University, Department of Physics, Guwahati, India
\item \Idef{org42}Helmholtz-Institut f\"{u}r Strahlen- und Kernphysik, Rheinische Friedrich-Wilhelms-Universit\"{a}t Bonn, Bonn, Germany
\item \Idef{org43}Helsinki Institute of Physics (HIP), Helsinki, Finland
\item \Idef{org44}High Energy Physics Group,  Universidad Aut\'{o}noma de Puebla, Puebla, Mexico
\item \Idef{org45}Hiroshima University, Hiroshima, Japan
\item \Idef{org46}Hochschule Worms, Zentrum  f\"{u}r Technologietransfer und Telekommunikation (ZTT), Worms, Germany
\item \Idef{org47}Horia Hulubei National Institute of Physics and Nuclear Engineering, Bucharest, Romania
\item \Idef{org48}Indian Institute of Technology Bombay (IIT), Mumbai, India
\item \Idef{org49}Indian Institute of Technology Indore, Indore, India
\item \Idef{org50}Indonesian Institute of Sciences, Jakarta, Indonesia
\item \Idef{org51}INFN, Laboratori Nazionali di Frascati, Frascati, Italy
\item \Idef{org52}INFN, Sezione di Bari, Bari, Italy
\item \Idef{org53}INFN, Sezione di Bologna, Bologna, Italy
\item \Idef{org54}INFN, Sezione di Cagliari, Cagliari, Italy
\item \Idef{org55}INFN, Sezione di Catania, Catania, Italy
\item \Idef{org56}INFN, Sezione di Padova, Padova, Italy
\item \Idef{org57}INFN, Sezione di Roma, Rome, Italy
\item \Idef{org58}INFN, Sezione di Torino, Turin, Italy
\item \Idef{org59}INFN, Sezione di Trieste, Trieste, Italy
\item \Idef{org60}Inha University, Incheon, Republic of Korea
\item \Idef{org61}Institut de Physique Nucl\'{e}aire d'Orsay (IPNO), Institut National de Physique Nucl\'{e}aire et de Physique des Particules (IN2P3/CNRS), Universit\'{e} de Paris-Sud, Universit\'{e} Paris-Saclay, Orsay, France
\item \Idef{org62}Institute for Nuclear Research, Academy of Sciences, Moscow, Russia
\item \Idef{org63}Institute for Subatomic Physics, Utrecht University/Nikhef, Utrecht, Netherlands
\item \Idef{org64}Institute for Theoretical and Experimental Physics, Moscow, Russia
\item \Idef{org65}Institute of Experimental Physics, Slovak Academy of Sciences, Ko\v{s}ice, Slovakia
\item \Idef{org66}Institute of Physics, Homi Bhabha National Institute, Bhubaneswar, India
\item \Idef{org67}Institute of Physics of the Czech Academy of Sciences, Prague, Czech Republic
\item \Idef{org68}Institute of Space Science (ISS), Bucharest, Romania
\item \Idef{org69}Institut f\"{u}r Kernphysik, Johann Wolfgang Goethe-Universit\"{a}t Frankfurt, Frankfurt, Germany
\item \Idef{org70}Instituto de Ciencias Nucleares, Universidad Nacional Aut\'{o}noma de M\'{e}xico, Mexico City, Mexico
\item \Idef{org71}Instituto de F\'{i}sica, Universidade Federal do Rio Grande do Sul (UFRGS), Porto Alegre, Brazil
\item \Idef{org72}Instituto de F\'{\i}sica, Universidad Nacional Aut\'{o}noma de M\'{e}xico, Mexico City, Mexico
\item \Idef{org73}iThemba LABS, National Research Foundation, Somerset West, South Africa
\item \Idef{org74}Johann-Wolfgang-Goethe Universit\"{a}t Frankfurt Institut f\"{u}r Informatik, Fachbereich Informatik und Mathematik, Frankfurt, Germany
\item \Idef{org75}Joint Institute for Nuclear Research (JINR), Dubna, Russia
\item \Idef{org76}Korea Institute of Science and Technology Information, Daejeon, Republic of Korea
\item \Idef{org77}KTO Karatay University, Konya, Turkey
\item \Idef{org78}Laboratoire de Physique Subatomique et de Cosmologie, Universit\'{e} Grenoble-Alpes, CNRS-IN2P3, Grenoble, France
\item \Idef{org79}Lawrence Berkeley National Laboratory, Berkeley, California, United States
\item \Idef{org80}Lund University Department of Physics, Division of Particle Physics, Lund, Sweden
\item \Idef{org81}Nagasaki Institute of Applied Science, Nagasaki, Japan
\item \Idef{org82}Nara Women{'}s University (NWU), Nara, Japan
\item \Idef{org83}National and Kapodistrian University of Athens, School of Science, Department of Physics , Athens, Greece
\item \Idef{org84}National Centre for Nuclear Research, Warsaw, Poland
\item \Idef{org85}National Institute of Science Education and Research, Homi Bhabha National Institute, Jatni, India
\item \Idef{org86}National Nuclear Research Center, Baku, Azerbaijan
\item \Idef{org87}National Research Centre Kurchatov Institute, Moscow, Russia
\item \Idef{org88}Niels Bohr Institute, University of Copenhagen, Copenhagen, Denmark
\item \Idef{org89}Nikhef, National institute for subatomic physics, Amsterdam, Netherlands
\item \Idef{org90}NRC Kurchatov Institute IHEP, Protvino, Russia
\item \Idef{org91}NRNU Moscow Engineering Physics Institute, Moscow, Russia
\item \Idef{org92}Nuclear Physics Group, STFC Daresbury Laboratory, Daresbury, United Kingdom
\item \Idef{org93}Nuclear Physics Institute of the Czech Academy of Sciences, \v{R}e\v{z} u Prahy, Czech Republic
\item \Idef{org94}Oak Ridge National Laboratory, Oak Ridge, Tennessee, United States
\item \Idef{org95}Ohio State University, Columbus, Ohio, United States
\item \Idef{org96}Petersburg Nuclear Physics Institute, Gatchina, Russia
\item \Idef{org97}Physics department, Faculty of science, University of Zagreb, Zagreb, Croatia
\item \Idef{org98}Physics Department, Panjab University, Chandigarh, India
\item \Idef{org99}Physics Department, University of Jammu, Jammu, India
\item \Idef{org100}Physics Department, University of Rajasthan, Jaipur, India
\item \Idef{org101}Physikalisches Institut, Eberhard-Karls-Universit\"{a}t T\"{u}bingen, T\"{u}bingen, Germany
\item \Idef{org102}Physikalisches Institut, Ruprecht-Karls-Universit\"{a}t Heidelberg, Heidelberg, Germany
\item \Idef{org103}Physik Department, Technische Universit\"{a}t M\"{u}nchen, Munich, Germany
\item \Idef{org104}Research Division and ExtreMe Matter Institute EMMI, GSI Helmholtzzentrum f\"ur Schwerionenforschung GmbH, Darmstadt, Germany
\item \Idef{org105}Rudjer Bo\v{s}kovi\'{c} Institute, Zagreb, Croatia
\item \Idef{org106}Russian Federal Nuclear Center (VNIIEF), Sarov, Russia
\item \Idef{org107}Saha Institute of Nuclear Physics, Homi Bhabha National Institute, Kolkata, India
\item \Idef{org108}School of Physics and Astronomy, University of Birmingham, Birmingham, United Kingdom
\item \Idef{org109}Secci\'{o}n F\'{\i}sica, Departamento de Ciencias, Pontificia Universidad Cat\'{o}lica del Per\'{u}, Lima, Peru
\item \Idef{org110}Shanghai Institute of Applied Physics, Shanghai, China
\item \Idef{org111}St. Petersburg State University, St. Petersburg, Russia
\item \Idef{org112}Stefan Meyer Institut f\"{u}r Subatomare Physik (SMI), Vienna, Austria
\item \Idef{org113}SUBATECH, IMT Atlantique, Universit\'{e} de Nantes, CNRS-IN2P3, Nantes, France
\item \Idef{org114}Suranaree University of Technology, Nakhon Ratchasima, Thailand
\item \Idef{org115}Technical University of Ko\v{s}ice, Ko\v{s}ice, Slovakia
\item \Idef{org116}Technische Universit\"{a}t M\"{u}nchen, Excellence Cluster 'Universe', Munich, Germany
\item \Idef{org117}The Henryk Niewodniczanski Institute of Nuclear Physics, Polish Academy of Sciences, Cracow, Poland
\item \Idef{org118}The University of Texas at Austin, Austin, Texas, United States
\item \Idef{org119}Universidad Aut\'{o}noma de Sinaloa, Culiac\'{a}n, Mexico
\item \Idef{org120}Universidade de S\~{a}o Paulo (USP), S\~{a}o Paulo, Brazil
\item \Idef{org121}Universidade Estadual de Campinas (UNICAMP), Campinas, Brazil
\item \Idef{org122}Universidade Federal do ABC, Santo Andre, Brazil
\item \Idef{org123}University College of Southeast Norway, Tonsberg, Norway
\item \Idef{org124}University of Cape Town, Cape Town, South Africa
\item \Idef{org125}University of Houston, Houston, Texas, United States
\item \Idef{org126}University of Jyv\"{a}skyl\"{a}, Jyv\"{a}skyl\"{a}, Finland
\item \Idef{org127}University of Liverpool, Liverpool, United Kingdom
\item \Idef{org128}University of Tennessee, Knoxville, Tennessee, United States
\item \Idef{org129}University of the Witwatersrand, Johannesburg, South Africa
\item \Idef{org130}University of Tokyo, Tokyo, Japan
\item \Idef{org131}University of Tsukuba, Tsukuba, Japan
\item \Idef{org132}Universit\'{e} Clermont Auvergne, CNRS/IN2P3, LPC, Clermont-Ferrand, France
\item \Idef{org133}Universit\'{e} de Lyon, Universit\'{e} Lyon 1, CNRS/IN2P3, IPN-Lyon, Villeurbanne, Lyon, France
\item \Idef{org134}Universit\'{e} de Strasbourg, CNRS, IPHC UMR 7178, F-67000 Strasbourg, France, Strasbourg, France
\item \Idef{org135} Universit\'{e} Paris-Saclay Centre d¿\'Etudes de Saclay (CEA), IRFU, Department de Physique Nucl\'{e}aire (DPhN), Saclay, France
\item \Idef{org136}Universit\`{a} degli Studi di Foggia, Foggia, Italy
\item \Idef{org137}Universit\`{a} degli Studi di Pavia, Pavia, Italy
\item \Idef{org138}Universit\`{a} di Brescia, Brescia, Italy
\item \Idef{org139}Variable Energy Cyclotron Centre, Homi Bhabha National Institute, Kolkata, India
\item \Idef{org140}Warsaw University of Technology, Warsaw, Poland
\item \Idef{org141}Wayne State University, Detroit, Michigan, United States
\item \Idef{org142}Westf\"{a}lische Wilhelms-Universit\"{a}t M\"{u}nster, Institut f\"{u}r Kernphysik, M\"{u}nster, Germany
\item \Idef{org143}Wigner Research Centre for Physics, Hungarian Academy of Sciences, Budapest, Hungary
\item \Idef{org144}Yale University, New Haven, Connecticut, United States
\item \Idef{org145}Yonsei University, Seoul, Republic of Korea
\end{Authlist}
\endgroup